\documentclass{jpsj-suppl}

\usepackage{txfonts}
\usepackage{graphicx}
\usepackage{amsmath}
\usepackage{amssymb}

\newcommand{\transpose}{\intercal}

\title{\vspace{-12pt}{\normalsize\rm \null \hfill ADP-15-45/T947~~~~\null}\\
$N^*$ Spectroscopy from Lattice QCD: The Roper Explained}

\author{
Derek \textsc{Leinweber}$^{1}$,
Waseem \textsc{Kamleh}$^{1}$,
Adrian \textsc{Kiratidis}$^{1}$,
Zhan-Wei \textsc{Liu}$^{1}$,
Selim \textsc{Mahbub}$^{1,2}$,
Dale \textsc{Roberts}$^{1,3}$,
Finn \textsc{Stokes}$^{1}$,
Anthony W. \textsc{Thomas}$^{1,4}$ and
Jiajun \textsc{Wu}$^{1}$
}

\inst{$^{1}$Special Research Centre for the Subatomic Structure
  of Matter (CSSM),\\Department of Physics, University of
  Adelaide, South Australia 5005, Australia\\[4pt]
$^{2}$CSIRO, Data61, 15 College Road, Sandy
  Bay, TAS 7005, Australia\\[4pt]
$^{3}$National Computational Infrastructure (NCI),\\Australian
  National University, Australian Capital Territory 0200, Australia\\[4pt]
$^{4}$ARC Centre of Excellence for Particle Physics at the Terascale
  (CoEPP),\\Department of Physics, University of Adelaide, South
  Australia 5005, Australia 
}

\email{derek.leinweber@adelaide.edu.au}

\recdate{November 11, 2015}

\abst{This brief review focuses on the low-lying even- and odd-parity
  excitations of the nucleon obtained in recent lattice QCD
  calculations.  Commencing with a survey of the 2014-15 literature
  we'll see that results for the first even-parity excitation energy
  can differ by as much as 1 GeV, a rather unsatisfactory situation.
  Following a brief review of the methods used to isolate excitations
  of the nucleon in lattice QCD, and drawing on recent advances, we'll
  see how a consensus on the low-lying spectrum has emerged among many
  different lattice groups.  To provide insight into the nature of
  these states we'll review the wave functions and electromagnetic
  form factors that are available for a few of these states.
  Consistent with the Luscher formalism for extracting phase shifts
  from finite volume spectra, the Hamiltonian approach to effective
  field theory in finite volume can provide guidance on the manner in
  which physical quantities manifest themselves in the finite volume
  of the lattice.  With this insight, we will address the question;
  Have we seen the Roper in lattice QCD?
}

\kword{lattice quantum chromodynamics, baryon spectrum, excited
  states, wave functions, form factors, finite volume}

\begin{document}
\maketitle

\section{Introduction}

\begin{figure}[t]
\centering
\includegraphics[width=0.4\hsize]{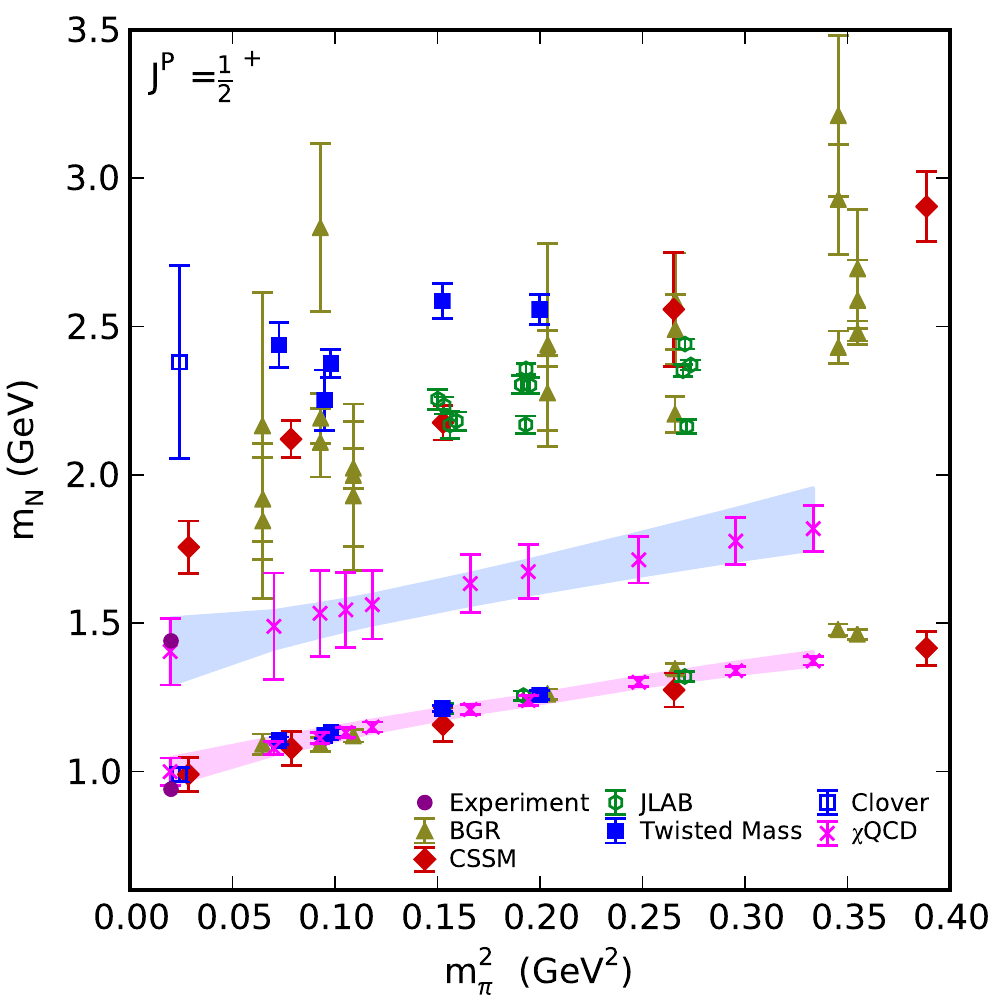} \quad
\includegraphics[width=0.53\hsize]{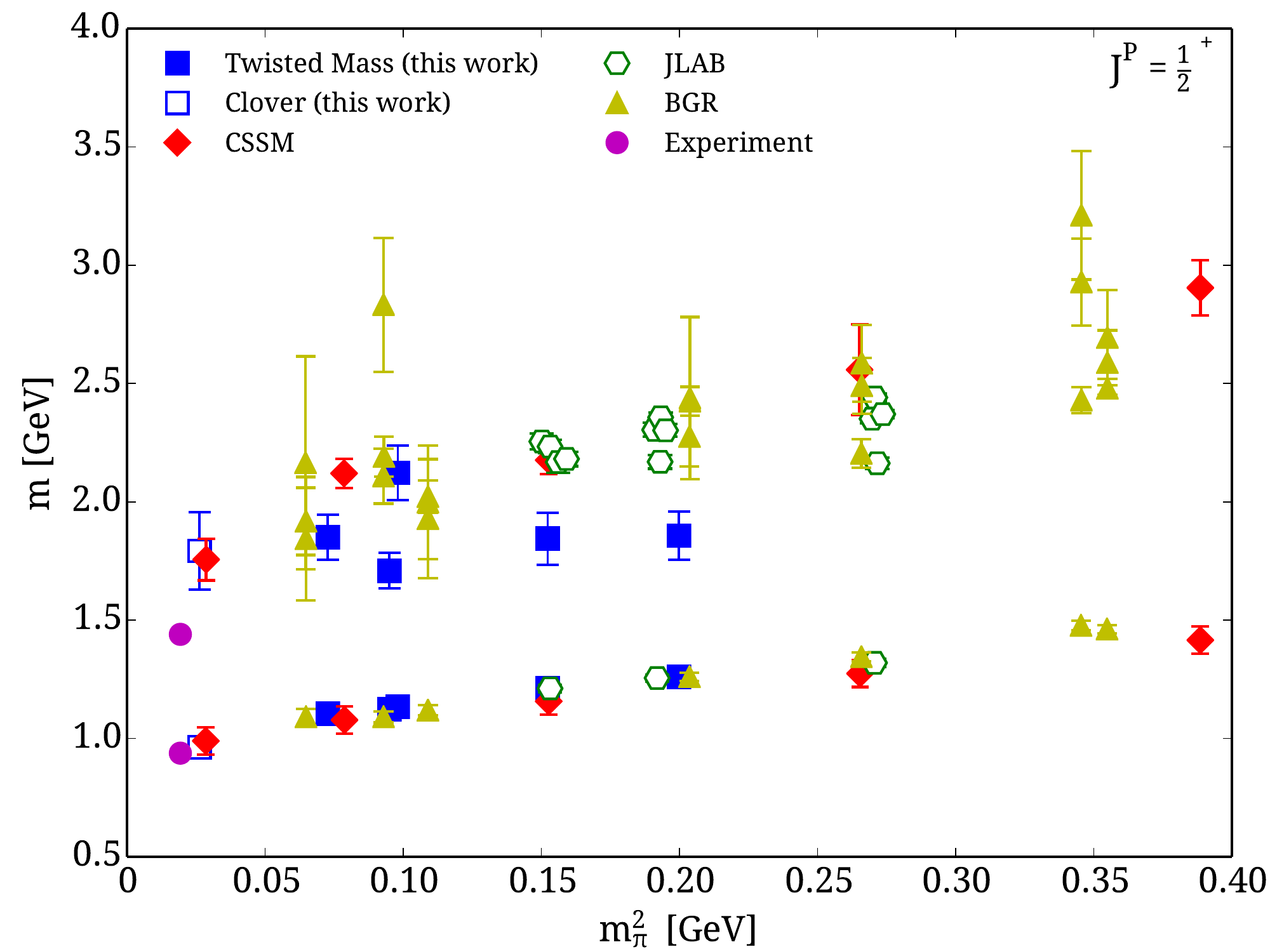}
\caption{The low-lying positive-parity spectrum of the nucleon
  observed in lattice QCD calculations circa January to March 2014.
  Figures are reproduced from the $\chi$QCD collaboration
  \cite{Liu:2014jua} (left) and the Cyprus group
  \cite{Alexandrou:2013fsu} (right).  Results from the
  Berlin-Graz-Regensburg (BGR) collaboration~\cite{Engel:2013ig},
  Centre for the Subatomic Structure of Matter
  (CSSM)~\cite{Mahbub:2010rm}, JLab Hadron Spectrum Collaboration
  (JLAB)~\cite{Edwards:2011jj}, $\chi$QCD collaboration
  \cite{Liu:2014jua} and Cyprus group (labelled Twisted Mass and
  Clover)~\cite{Alexandrou:2013fsu} are presented.
}
\label{fig:chiqcdCyprus}
\end{figure}

In this brief review, we survey recent results for the
nucleon spectrum obtained from numerical simulations of lattice QCD.
We begin by reviewing the status of the nucleon spectrum as reported
by the $\chi$QCD collaboration \cite{Liu:2014jua} and the Cyprus group
\cite{Alexandrou:2013fsu} in early 2014.  Focusing first on the
left-hand plot of Fig.~\ref{fig:chiqcdCyprus}, at $m_\pi^2 \sim 0.15\,{\rm
  GeV}^2$ estimates for the mass of the first even parity excitation
of the nucleon range from 1.5 GeV by the $\chi$QCD collaboration
\cite{Liu:2014jua}  to 2.5 GeV by the Cyprus group
\cite{Alexandrou:2013fsu}. 

In comparing the two plots of Fig.~\ref{fig:chiqcdCyprus}, aside from
the Cyprus group not including the $\chi$QCD collaboration results,
the most notable change is a systematic shift of the Cyprus group's
results from values $\sim 2.5$ GeV to $\sim 1.8$ GeV, a shift much
larger than the statistical uncertainties.  Cyprus results in the
left-hand plot are from version 1 of the arXiv preprint of
Ref.~\cite{Alexandrou:2013fsu} from February 2013 whereas the
right-hand plot illustrates the results published in
Ref.~\cite{Alexandrou:2013fsu} a year later.  
These results were subsequently improved in the analysis of
Ref.~\cite{Alexandrou:2014mka} by the Cyprus group.  It is these final
results that have enabled a consensus to form.
In the following we will examine the physics behind the evolution of
lattice QCD results towards this consensus.  However, we will begin
with a brief review of the CSSM and JLab Hadron Spectrum Collaboration
(HSC) results.

\section{CSSM and JLab HSC results}

In the CSSM approach, a basis of nucleon interpolating fields is
created from the three local nucleon interpolating fields with various
levels of gauge invariant Gaussian smearing \cite{Gusken:1989qx}
applied to the spatial dimensions of the quark sources and sinks.  The
superpositions of excited states are dependent on the level of
smearing applied \cite{Mahbub:2010rm}, such that linear combinations
of the smeared sources are constructed to isolate the states of the
spectrum through the generalised eigenvalue problem.  The outcome of
the calculation is a superposition of Gaussians of various widths.
Indeed the pattern of the superposition strengths contained in the
eigenvectors is precisely that required to create nodes in the excited
state wave functions \cite{Mahbub:2013ala}.

In the JLab HSC approach \cite{Edwards:2011jj}, a basis of
interpolating operators with good total angular momentum in the
continuum is developed.  These are then subduced to the various
lattice irreducible representations by creating lattice derivative
operators respecting the cubic symmetry of the lattice.  They find
that the subduced operators retain a memory of their continuum
antecedents to a remarkable degree.  While their approach resembles
that of a single level of smearing in the quark sources and sinks, the
extended nature of their operators aids in isolating the eigenstates
of the spectrum.

\begin{figure}[t]
\centering
\includegraphics[angle=90,width=0.49\hsize]{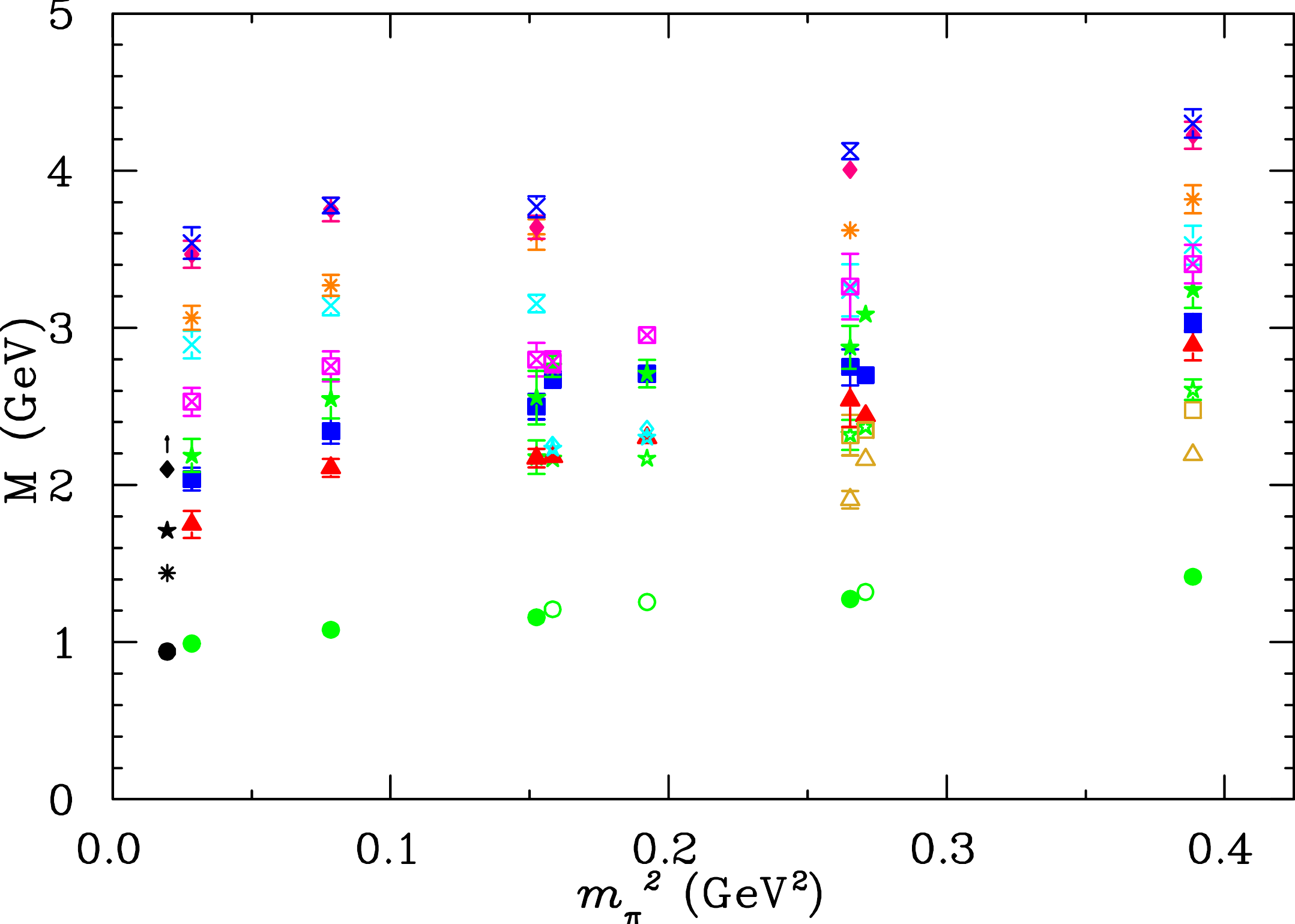}
\includegraphics[angle=90,width=0.49\hsize]{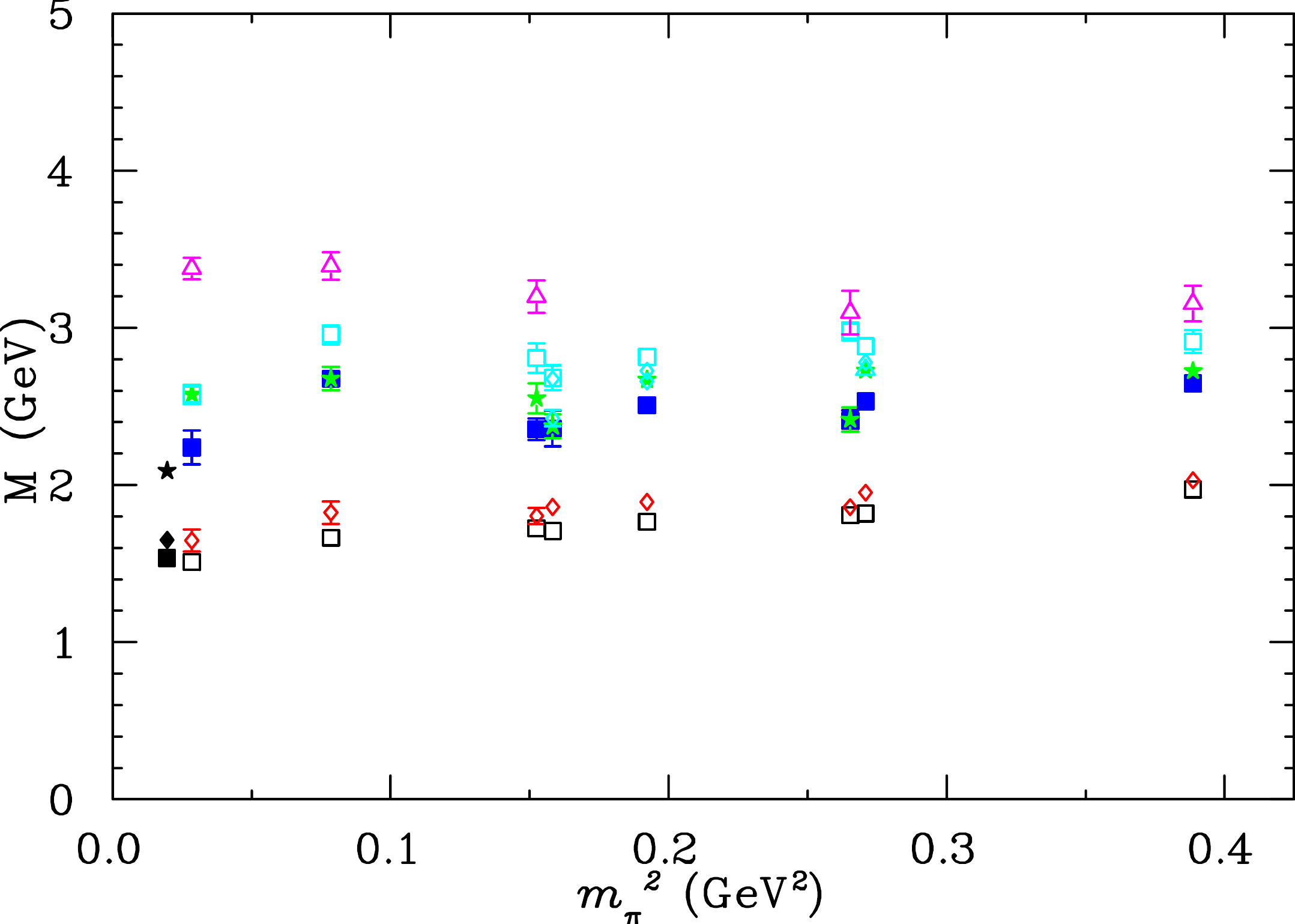}
\caption{Comparison of CSSM \cite{Mahbub:2010rm} and JLab HSC
  \cite{Edwards:2011jj} results for the low-lying even-parity (left)
  and odd-parity (right) nucleon spectrum.  The three pion masses
  corresponding to the HSC collaboration are easily identified by the
  open circles plotted for the ground state nucleon in the left-hand
  plot.
  \vspace{-0.6cm}
}
\label{fig:CSSMJLab}
\end{figure}

Fig.~\ref{fig:CSSMJLab} presents a comparison of CSSM and JLab HSC
results made as the HSC results became available.  Despite forming the
correlation matrix in completely different manners, the qualitative
agreement is remarkable.  One must remember that the spatial volumes
of the simulations are different and therefore the multi-particle
scattering thresholds are different.  The CSSM results have a volume
with side $\sim 3$ fm whereas the HSC results are from a smaller
volume of $\sim 2$ fm on a side.

Indeed the most notable discrepancy in Fig.~\ref{fig:CSSMJLab} is the
first even-parity excitation at $m_\pi^2 \simeq 0.27\,{\rm GeV}^2$
observed by the CSSM.  However, this is attributed to the volume
dependence of the noninteracting $P$-wave $N \pi$ scattering threshold
given by $E_\pi + E_N$ with $E_i = \sqrt{p^2 + m_i^2}$ and the
momentum $p$ governed by the lowest nontrivial momentum on the
lattice, $p = 2\pi/L$, where $L$ is the length of the longest spatial
dimension.  The CSSM point lies just below this threshold.  The quantitative
agreement between all other states is surprising.  The agreement at
$m_\pi^2 \simeq 0.17\,{\rm GeV}^2$ for the first even-parity excitation
is particularly important in the discussion that follows.

\section{$\chi$QCD Collaboration Results}

To better understand the origin of the discrepancy between the
$\chi$QCD collaboration results and the CSSM/HSC results depicted in the
left-hand plot of Fig.~\ref{fig:chiqcdCyprus}, the $\chi$QCD
collaboration performed an important controlled experiment
\cite{Liu:2014jua}.  They analysed the gauge fields and quark
propagators created by the HSC, thus eliminating all systematic
uncertainties other than the final analysis of the baryon correlator.
The lightest of the three HSC ensembles were considered where $m_\pi^2
\simeq 0.17\,{\rm GeV}^2$.  In the usual tradition, they analysed the
proton correlation function generated from $\chi_1(x)\,
\overline\chi_1(0)$ where $\chi_1(x) = \varepsilon^{abc} (\,
u^{\transpose a}(x) \, C\gamma_5 \, d^b(x)\, )\, u^c(x)$ is the
traditional local nucleon interpolating field.  Using their Sequential
Empirical Bayesian analysis, they resolved the mass of the first
even-parity excitation of the nucleon to be 300 MeV below that
obtained by the HSC collaboration \cite{Liu:2014jua}.  

\begin{figure}[t]
\centering
\includegraphics[clip=true,trim=2.9cm 0.0cm 2.9cm 0.0cm,width=0.45\hsize]{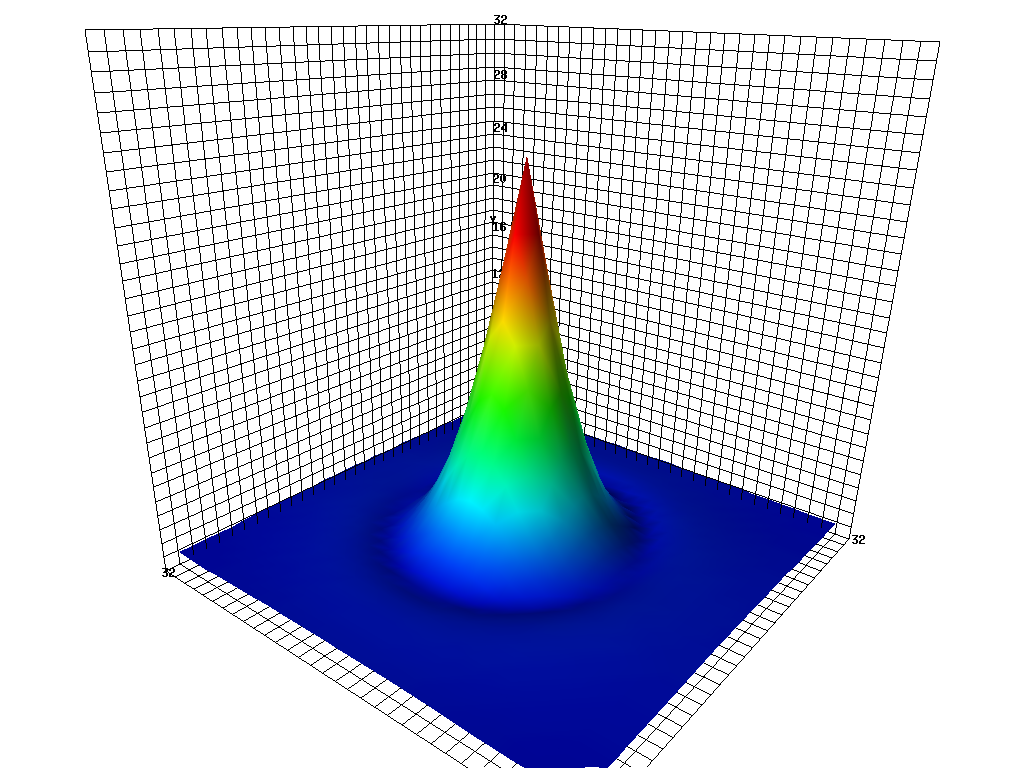}
\includegraphics[clip=true,trim=2.9cm 0.0cm 2.9cm 0.0cm,width=0.45\hsize]{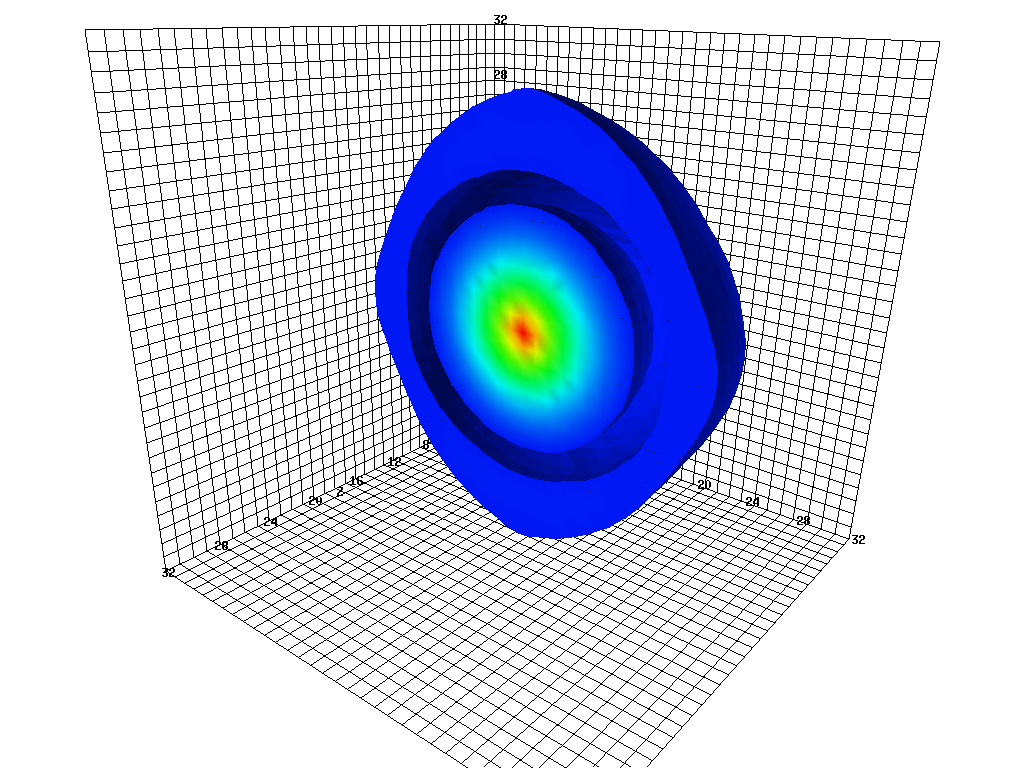}
\caption{The $d$-quark probability distribution in the first even
  parity excitation of the nucleon about two $u$ quarks fixed at the
  origin at the centre of the plots
  \cite{Roberts:2013ipa,Roberts:2013oea}.  The CSSM calculations are
  at the lightest of quark masses considered providing $m_\pi = 156$
  MeV \cite{Aoki:2008sm}.  The isovolume threshold for rendering the
  probability distribution in the right-hand plot is $3.0 \times
  10^{-5}$.  The node occurs at 0.83 fm from the centre.
  The three-dimensional axis grid provides an indication of the
  positions of the $32^3$ lattice sites for the isovolume.  The
  PACS-CS scheme for determining the lattice spacing provides $a =
  0.0907(13)$ fm.
  \vspace{-0.6cm}
}
\label{fig:RoperNode}
\end{figure}

In accounting for this discrepancy, the $\chi$QCD collaboration pointed
to the small size of the HSC collaboration's quark-propagator source
arguing that they would not be sensitive to the node in the wave
function of the first even-parity excitation.
Fig.~\ref{fig:RoperNode} illustrates the node structure observed by
the CSSM collaboration.  The $\chi$QCD collaboration argued
that the basis of interpolators considered by the HSC collaboration
was insufficient to span the space, thus producing an erroneously
large excitation energy.  

However, this is not the case.  As illustrated in
Figs.~\ref{fig:chiqcdCyprus} and \ref{fig:CSSMJLab}, the CSSM result
for the first excited state is in excellent agreement with the HSC
result at $m_\pi^2 \simeq 0.17\,{\rm GeV}^2$.  A key feature of the
CSSM approach is to consider unusually large fermion source sizes to
ensure the correlation matrix basis is sensitive to the node and does
indeed effectively span the space of relevant basis states for the
low-lying spectrum.  This was explored extensively in
Ref.~\cite{Mahbub:2010rm} where RMS smearing radii from 0.1 to 1.5 fm
were investigated, extending well beyond the node at 0.83 fm.  

We will return to this issue again after discussing the Athens Model
Independent Analysis Scheme employed by the Cyprus collaboration.

\section{Athens Model Independent Analysis Scheme}

The Athens Model Independent Analysis Scheme (AMIAS) explored by the
Cyprus group \cite{Alexandrou:2014mka} is a robust and complementary
approach to the analysis of hadron correlation functions.  Its
strength lies in that it does not require human intervention in
identifying plateaus in effective mass plots.  Indeed it has been key
is revising the Cyprus' group's earlier nucleon spectrum results
\cite{Alexandrou:2013fsu} and thus forming a consensus among the
Cyprus, JLab and CSSM collaborations on the low-lying excitation
spectrum of the nucleon.

The AMIAS method works with the same matrix of correlation functions
considered in the standard generalised eigenvalue approach.  It
exploits the small time separations in the correlation functions where
the excited states contribute strongly and statistical errors are
small.  One considers the standard spectral decomposition of the
correlation matrix
\begin{equation}
G_{ij}(t) = \sum_{\alpha=0}^{N_{\rm states}} \,  A^\alpha_i \, A^{\dagger \alpha}_j \, e^{-E_\alpha\, t} \, . \quad
  i,j = 1,\ldots, N_{\rm interpolators} \, ,
\end{equation}
where $E_\alpha$ is the energy of eigenstate $\alpha$ (increasing with
increasing $\alpha$) and the amplitudes $A^\alpha_i$ and $A^{\dagger
  \alpha}_j$ indicate the overlap with the $j$'th source interpolating
and $i$'th sink interpolating field with state $\alpha$.

However, what sets this approach apart from other approaches is that
it uses Monte Carlo-based importance sampling to determine the
parameters of the spectral decomposition, $A^\alpha_i$ and $E_\alpha$.
Using a parallel-tempering algorithm to avoid local-minima traps
\cite{Alexandrou:2014mka}, values are proposed and selected with the
probability $\exp(-\chi^2/2)$, governed by the $\chi^2$ of the fit.
Final parameter values and uncertainties are determined by fitting a
Gaussian to their probability distributions. The number of states
required in the fit, $N_{\rm states}$, is determined by increasing the
number of states until the highest states show no preference in their
fit values.

\begin{figure}[t]
\centering
\includegraphics[width=0.55\hsize]{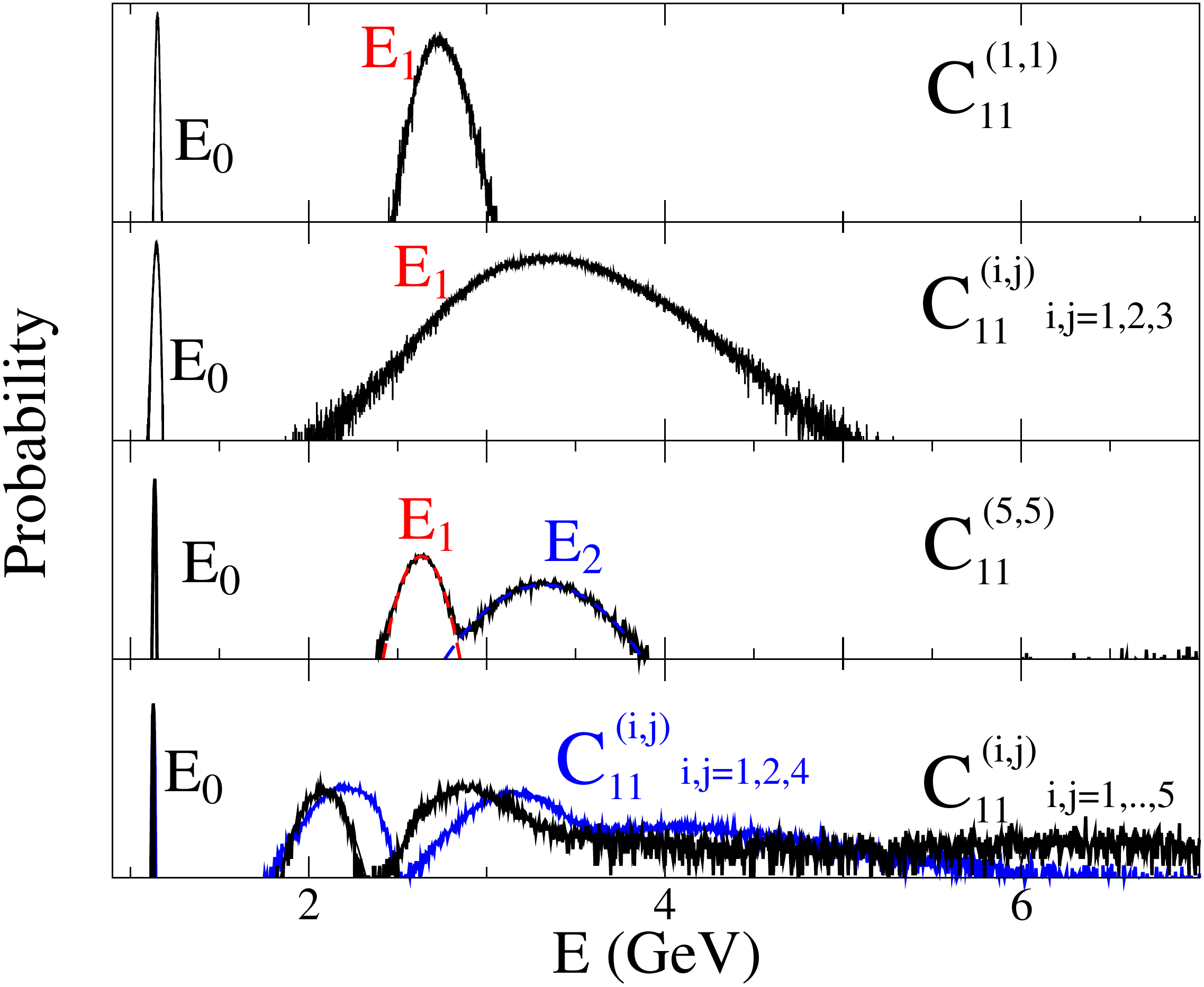}
\caption{Reproduced from the Cyprus' group's work investigating the
  Athens Model Independent Analysis Scheme (AMIAS)
  \cite{Alexandrou:2014mka} this plot illustrates the positive parity
  nucleon spectrum with various subsets of their $5\times 5$
  correlation matrix obtained from the $\chi_1(x)\,
  \overline\chi_1(0)$ correlator with increasing levels of Gaussian
  smearing applied in the quark sources and sinks.  The results in
  black in the bottom panel of the figure indicate the best results
  obtained from an analysis of the entire correlation matrix, showing
  a first excitation at $\sim 2.1$ GeV relative to a ground state
  nucleon at $\sim 1.2$ GeV.  The top panel illustrates results
  obtained from a single correlator with a small amount of smearing.
  The second panel illustrates results from a subset of the
  correlation matrix where the smearing of the interpolating fields
  is small and insensitive to the node in the wave function of the
  first excitation.  Similar to the top panel, the third panel
  illustrates single correlator results, but this time with a large
  smeared source considered.  The blue curve in the bottom panel omits
  the third and fifth (largest) of the smeared source/sink
  correlators.
  \vspace{-0.6cm}
}
\label{fig:AMIAS}
\end{figure}

Figure \ref{fig:AMIAS} illustrates the key results of
Ref.~\cite{Alexandrou:2014mka}.  In the smeared-operator correlation
matrix approach, the combination of both large and small smearing
extents is critical to obtaining the correct first positive-parity
excitation of the nucleon.  One cannot get the correct result from the
analysis of a single correlator, even when the source is large enough
to span the node of the first excitation.  There is insufficient
information in a single correlation function to correctly reproduce
the spectrum observed when analysing all $5\times5 = 25$ correlation
functions.

This then raises a concern in the Sequential Empirical Bayesian
analysis which focuses on a single correlation function.  With
insufficient information in the single correlator, one becomes
concerned about the role of the algorithm in producing a low-lying
result for the first positive-parity excitation of the nucleon.

The CSSM collaboration has had a first look at the electromagnetic
form factors of the lowest lying even- and odd-parity nucleon
excitations \cite{Owen:2013pfa,Owen:2014txa}.  The electric form
factors \cite{Owen:2013pfa} confirm a significantly larger
distribution of the quark flavours in these states.  Remarkably at
heavier quark masses, the first even parity excitations produce
magnetic moments similar to the ground state nucleons, consistent with
the idea of a $2s$-state excitation \cite{Owen:2014txa}.  In the odd
parity sector, the lowest lying excitation has magnetic moments
consistent with quark model expectations \cite{Owen:2014txa}.

\section{A consensus is established}

Having ascertained the necessity of analysing a matrix of correlation
functions constructed with a variety of sources sensitive to the node
structures of the excited-state wave functions, we are now able to
update the results of Fig.~\ref{fig:chiqcdCyprus}.  

In doing so the CSSM collaboration has updated the statistics of the
results published in Refs.~\cite{Mahbub:2013ala} and
\cite{Mahbub:2012ri} to approximately 29,500 propagators on the
PACS-CS configurations \cite{Aoki:2008sm}.  Recall, these lattices
have a spatial length $L \simeq 3.0$ fm.
The Cyprus group's results \cite{Alexandrou:2014mka} are based on
twisted-mass fermion simulations and Wilson-clover simulations from
the QCDSF collaboration \cite{Bali:2012qs} at the lightest quark mass
considered.  Their volumes have length $L \simeq 2.8$ fm.  Complete
details are provided in Ref.~\cite{Alexandrou:2014mka}.
The JLab HSC results \cite{Edwards:2011jj} are based on anisotropic
Wilson-clover lattices with length $L \simeq 2$ fm.

\begin{figure}[t]
\centering
\includegraphics[width=0.49\hsize]{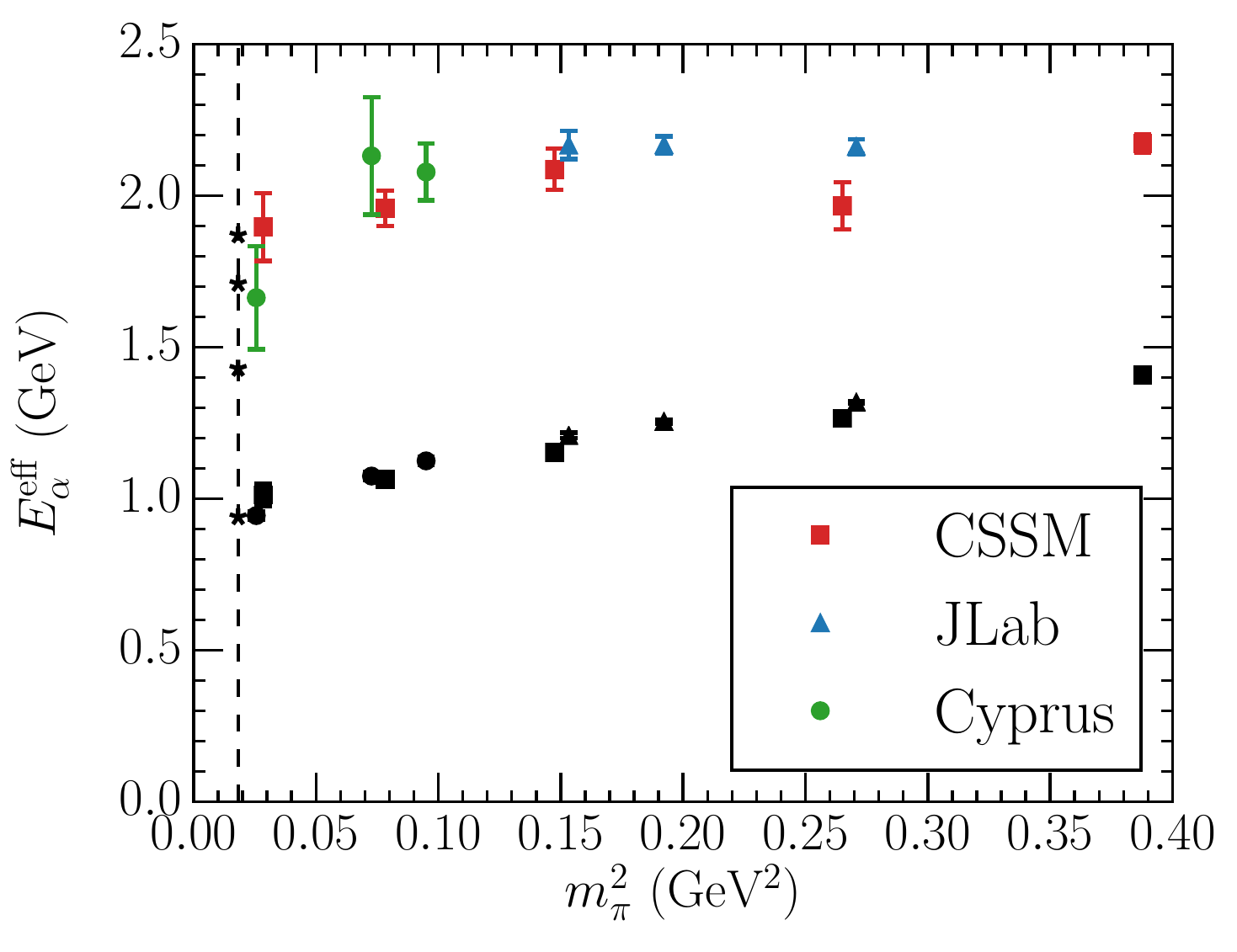}
\includegraphics[width=0.49\hsize]{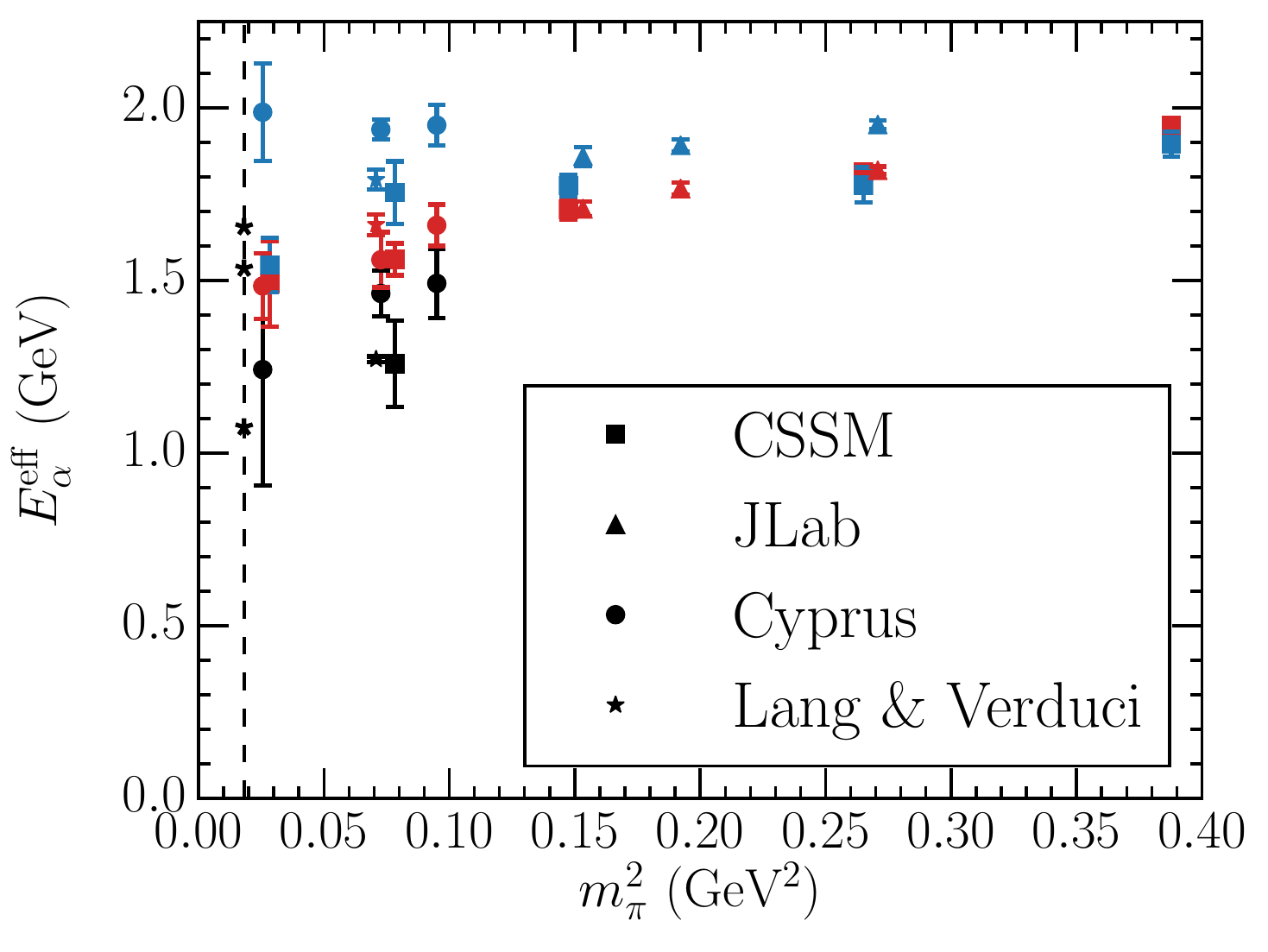}
\caption{Current results for the positive-parity (left) and
  negative-parity (right) nucleon spectrum as reported by the
  collaborations CSSM \cite{Mahbub:2013ala,Kiratidis:2015vpa} and
  herein, Cyprus \cite{Alexandrou:2014mka}, JLab \cite{Edwards:2011jj}
  and Lang and Verduci \cite{Lang:2012db}.  Results for the first two
  even-parity states of the nucleon are illustrated, while the first
  three states (identified by the colour of the plot symbols) are
  presented in the odd-parity channel.  Experimental results are
  illustrated on the vertical dashed line at the physical pion mass.
  A consensus on the low-lying spectrum has emerged among these
  lattice groups.
  \vspace{-0.6cm}
}
\label{fig:currentNucleonSpectrum}
\end{figure}

Figure~\ref{fig:currentNucleonSpectrum} presents results for the
ground and first even-parity excitation of the nucleon, and the first
three low-lying odd-parity states observed in lattice QCD
calculations.
In the odd-parity sector, new results from Lang and Verduci
\cite{Lang:2012db} are presented alongside new CSSM results
\cite{Kiratidis:2015vpa} demonstrating the robust nature of CSSM
analysis methods.  Both of these results investigate new interpolating
fields for exciting the $N^*$ spectrum.  

In the CSSM approach local five-quark operators resembling $N\pi$
states are considered.  While the total momentum of the $N\pi$ system
is constrained to zero, all possible momenta between the nucleon and
pion are allowed.  In contrast, Lang and Verduci
\cite{Lang:2012db} project the momenta of both the nucleon and pion to
zero, creating very strong overlap with the lowest lying $N\pi$
scattering state in the finite volume of the lattice.  

Both the JLab HSC \cite{Edwards:2011jj} and CSSM \cite{Mahbub:2012ri}
collaborations were able to resolve two low-lying resonant-like states near
the $N(1535)$ and $N(1650)$ of nature.  They were able to do so without
accessing the lowest-lying $N\pi$ scattering state.  However, Lang and
Verduci found it necessary to include the momentum projected $N\pi$
interpolating field with their analysis techniques \cite{Lang:2012db}.
While some have argued that the $N\pi$ scattering state must be
observed to get the rest of the spectrum correct, this is not the
case \cite{Kiratidis:2015vpa}.  Inclusion of the $N\pi$ scattering
state is sufficient \cite{Lang:2012db}, but not necessary
\cite{Kiratidis:2015vpa}.

\section{Have we seen the Roper?}

\begin{figure}[t]
\centering
\includegraphics[width=0.80\hsize]{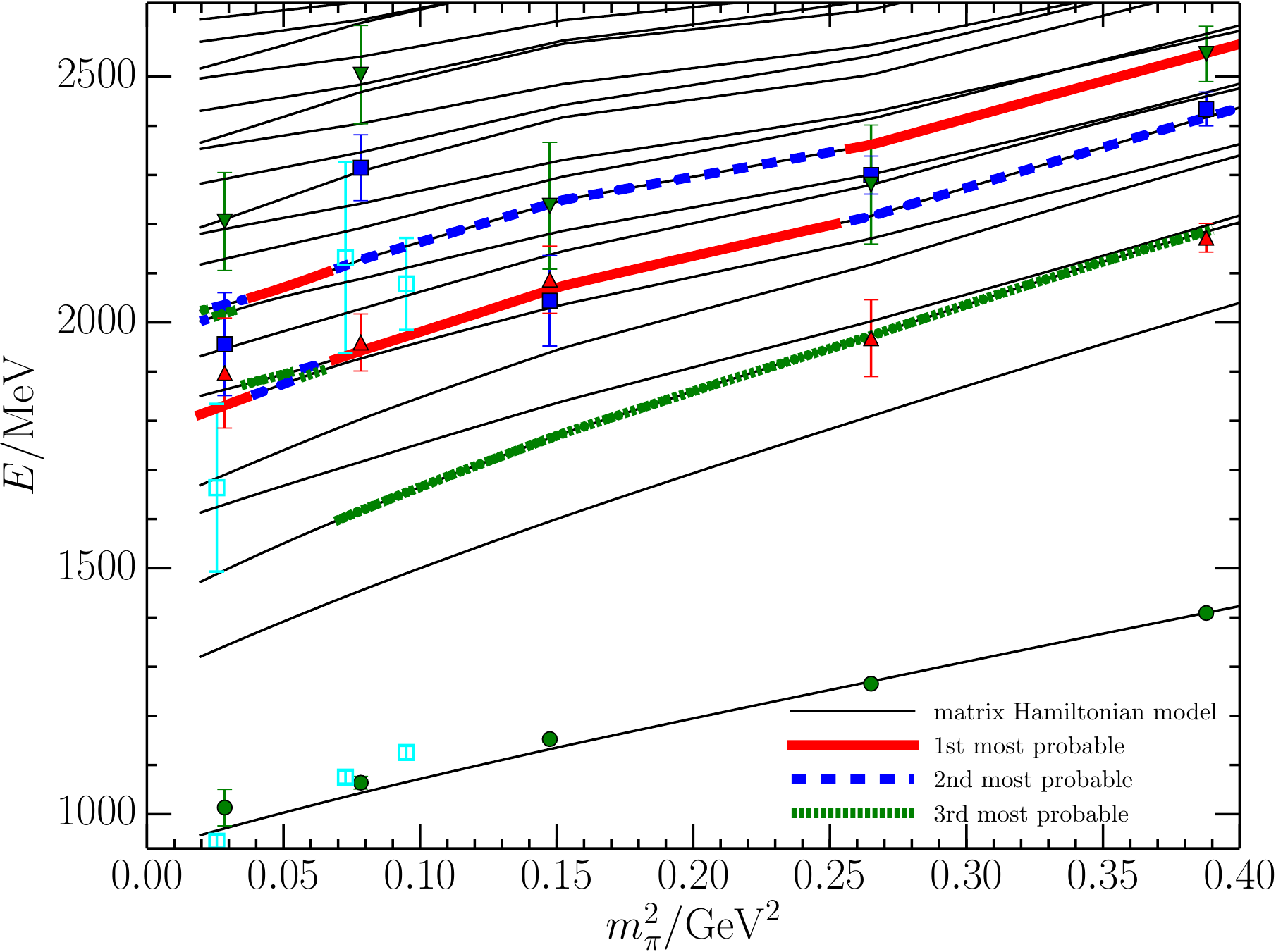}
\caption{Finite-volume energy eigenstates for the positive-parity
  nucleon spectrum from Hamiltonian Effective Field Theory
  \cite{Liu:2015} are compared with CSSM and Cyprus collaborations,
  the latter indicated in cyan.  These finite-volume states with $L
  \sim 3$ fm correspond to a dynamically-generated infinite-volume
  Roper-resonance pole position in agreement with the particle data
  tables.  States having a significant bare state component -- and
  thus more likely to be excited by three-quark operators -- are
  labelled ``most probable'' and marked solid red (4 to 5\%), dashed
  blue (3 to 4\%) and dotted green ($\sim 1$\%), where the percentage
  values indicate the typical bare mass contribution to the
  eigenstate.
  \vspace{-0.6cm}
}
\label{fig:Roper}
\end{figure}

In answering this question, one must have a formalism for connecting
the finite-volume spectrum of lattice QCD to the infinite-volume {\em
  resonances} of nature.  The CSSM collaboration has been
investigating Hamiltonian Effective Field Theory (HEFT)
\cite{Hall:2013qba} to make this connection.  Similar to the original
idea of Luscher \cite{Luscher:1985dn,Luscher:1986pf}, 
HEFT provides a robust link between the energy levels observed in the
finite volume of lattice QCD and the $Q^2$ dependence of the
scattering phase shift.  In this way, the CSSM \cite{Liu:2015} has
drawn on experimental data for the scattering phase shift,
inelasticity and pole position of the lowest-lying $J^P=1/2^+$ nucleon
resonance, the Roper resonance \cite{Roper:1964zza}, and used HEFT to
predict the positions of the finite-volume energy levels to be
observed in lattice QCD simulations in volumes of $\sim 3$ fm.

The lattice simulation results from the CSSM and Cyprus collaborations
employ three-quark operators to excite the spectrum.  For three-quark
operators, the couplings to two-particle dominated scattering states
are volume suppressed by $1/L^3$.  As both collaborations consider
relatively large lattice volumes with $L \sim 3$ fm, the HEFT states
having the largest contribution from the bare basis state are the
states more likely to be observed in the lattice QCD simulations.  For
example, in the ground state nucleon, the bare state is the dominant
contribution, providing $\sim 80$\% of the basis-state contributions.

Figure~\ref{fig:Roper} illustrates the positions of the energies from
the HEFT calculation with one bare mass coupled to $N\pi$, $N\sigma$,
and $\Delta\pi$ intermediate states \cite{Liu:2015}.  The colour
coding of the lines denotes the excited states having the largest
bare-state component and thus the most likely states to be excited in
the CSSM and Cyprus simulations.  Agreement between HEFT predictions
and lattice QCD results is observed as the lattice QCD results tend to
sit on the finite volume energy eigenstates dominated by the bare
state component.  At the lightest quark masses, most of the bare-state
strength lies in the first and second most probable states to be seen
in the lattice QCD results, having four to five times the bare-state
strength of the third most-probable state.  This explains why
low-lying scattering states are not seen in lattice QCD at light
quark masses.

The lower-lying states of the spectrum are dominated by the nearby
noninteracting basis states.  Just as the low-lying $\Lambda(1405)$
was found to be a $\overline K N$ molecular bound state
\cite{Menadue:2011pd,Hall:2014uca,Hall:2014gqa}, there is no low-lying
state in the regime of the Roper resonance with a significant bare
state component.
In order of increasing energy, the excited states are dominated by
$N\sigma(p=0)$, $N\pi(p=1)$, $N\sigma(p=0)$ and a mix of $N\pi(p=1)$
and $\Delta\pi(p=0)$ for the fourth excitation.  Here, the momenta in
parentheses indicate the back-to-back momenta of the meson and baryon
in units of $2\pi/L$.  Future lattice QCD simulations will aim to
excite these scattering states and accurately determine their
energies.  In this way one can constrain the HEFT from the first
principles of QCD and examine the manner in which QCD gives rise to
the resonance properties of the Roper excitation.

On the basis of our present analysis we conclude that we have seen at
least part of the the Roper in lattice QCD, the part looking most like
constituent quark model expectations \cite{Roberts:2013oea}.  The HEFT
model describes the Roper resonance of nature as a dynamically
generated resonance composed of only a small bare-state component,
observed in the region of the finite-volume lattice QCD results.

This research is supported by the National Computational Merit
Allocation Scheme and the Australian Research Council through grants
DP150103164, DP14010306, DP120104627 and LE120100181.


\providecommand{\href}[2]{#2}\begingroup\raggedright\endgroup

\end{document}